# Topic Diffusion Discovery based on Sparseness-constrained Non-negative Matrix Factorization

Yihuang Kang
*National Sun Yat-sen University*
*ykang@mis.nsysu.edu.tw*

Keng-Pei Lin
*National Sun Yat-sen University*
*kplin@mis.nsysu.edu.tw*

I-Ling Cheng
*National Sun Yat-sen University*
*chengi428@mis.nsysu.edu.tw*

**Abstract**

*Due to recent explosion of text data, researchers have been overwhelmed by ever-increasing volume of articles produced by different research communities. Various scholarly search websites, citation recommendation engines, and research databases have been created to simplify the text search tasks. However, it is still difficult for researchers to be able to identify potential research topics without doing intensive reviews on a tremendous number of articles published by journals, conferences, meetings, and workshops. In this paper, we consider a novel topic diffusion discovery technique that incorporates sparseness-constrained Non-negative Matrix Factorization with generalized Jensen-Shannon divergence to help understand term-topic evolutions and identify topic diffusions. Our experimental result shows that this approach can extract more prominent topics from large article databases, visualize relationships between terms of interest and abstract topics, and further help researchers understand whether given terms/topics have been widely explored or whether new topics are emerging from literature.*

## 1. Introduction

Information and Communication Technology (ICT) have been reshaping nearly every corner of the world, especially the way of people's communication. Today, our knowledge can be easily collected, stored, and distributed digitally. Researchers have been using new communication tools to facilitate the exchange of information. However, these document (text) data have grown exponentially in recent decades, and numerous research articles are created, digitized, and stored in various research databases. It is almost impossible for a researcher to do thoroughly literature search and review on a tremendous number of articles in order to understand recent progress of his/her research of interest. Many research databases, scholarly search engines, and citation recommendation systems have been created to help tackle this problem. To gather literature related to a research is much easier today, but we are still facing challenges to determine whether a research topic (or just a term) is rarely explored by others, just emerging from discussions of small groups, or potentially connected to topics in other fields of studies—problems of *topic diffusion discovery*.

In recent years, cheaper cost of computing power makes it possible to analyze vast amounts of textual data and to extract useful information from them within reasonable times. To cope with the aforementioned problems from large text, Topic Models [1] and related techniques are proposed to identify hidden semantic structures that may help us annotate, re-organize, and understand the contents of texts from higher-level abstract *topics*. The most well-known topic modeling techniques are *Latent Dirichlet Allocation* (LDA) [2] and *Non-negative Matrix Factorization* (NMF) [3]. LDA is a kind of generative probabilistic model that assumes documents exhibit multiple abstract topics consisting of some words/terms in a given set of vocabulary, whereas NMF is a sort of non-probabilistic dimensionality reduction technique that considers documents can be represented by additive combinations of some major parts (topics) of objects. Both approaches and their extensions have been successfully applied to topic modeling and other fields, but researchers have found that both are difficult to be used in dynamic or interactive fashions to track the diffusions of existing topics or to detect the emergences of new topics [4–7]. That is, many real-world topic modeling applications require existing topic models to be able to efficiently update topic models when new documents/text data stream appears, and to identify significant topic changes as text data increases over time.

In this paper, we introduce a novel topic diffusion discovery technique that combines a modified sparseness-constrained NMF [8] and *generalized Jensen-Shannon divergence* ($D_{GJS}$) [9–10] aiming at building topic models with more prominent topics and detecting topic diffusion by monitoring the changes of probability distributions of given terms among multiple topics. That is, for example, suppose we have built topic models from thousands of news articles. We might find that the probability distributions of "cell phone" and "camera" together in some topics (e.g. "Electronics" and "Lifestyle") have been changing dramatically in the past few decades, as cell phones with cameras have been transforming our mobile lifestyle in recent years—the *topic diffusion*. The goal of the proposed approach is to exhibit probability distributions of given terms associated with the topics (identified by aforementioned sparseness-constrained NMF) and evaluate its convergence or divergence of these probability distributions in different times. By "divergence" here, we mean the *Information Divergence* commonly used in measuring abrupt or evolutionary changes among two or multiple probability distributions. We here consider applying $D_{GJS}$ to the evaluation because of its boundedness property that provides certain limits of the divergence and statistically significant threshold that simplifies the evaluation of magnitude of the divergence.

The rest of this paper is organized as follows. In Section 2, we review related works of topic modeling and text analytics. NMF and its extensions, such as sparseness-constrained NMF techniques, are also discussed. We consider proposed approach in Section 3. In Section 4, we present and discuss our experimental results on a large number of research articles from a real-world research article database related to "Machine Learning". In Section 5, we conclude and summarize our findings.

## 2. Background and Related Work

Today, our collected knowledge can be easily digitized and stored into text data. The recent advance of ICT has facilitated the communication and processing of information, but has also resulted in the explosion of digital text data. Many researchers in text analytics communities have proposed techniques that automate the processing of text in order to extract information and knowledge from large amounts of unstructured text. *Topic modeling* [11] is one of the famous techniques that learns latent/hidden semantic structures of text consisting of multiple terms as abstract topics. In the last decade, the most popular probabilistic topic model was *Latent Dirichlet Allocation* (LDA) [2–3] which assumes that a document is composed of multiple topics and a topic is a distribution over a fixed set of pre-defined words (i.e. a general or domain-specific dictionary). LDA has been very successful because of its flexible models and interpretable topics. It has also been widely applied to various domains and many extensions have been proposed [5], [13]. Nevertheless, due to its probabilistic nature and sampling-based procedure, typical LDA might give inconsistent results from multiple runs and significant topic member (term) changes when text data does not exhibit clear topics [7], which hinders itself from real-world online or interactive text analytics applications. Instead, we here consider non-probabilistic topic models, *Nonnegative Matrix Factorization* (NMF) [3], which is a deterministic topic learning algorithm that could generates consistent results given the same text corpus and learning parameters, such as multiple runs with the same matrix initialization factor [14].

Similar to other matrix factorization techniques, such as *Singular Value Decomposition* (SVD) [15], NMF gives low-rank matrix approximation but more intuitive, parts-based, and additive representations of the original data matrix. It has been applied to various fields, such as classification/comparison of face recognitions, music transcription of signal processing, and topic modeling in text mining [3], [16], [17]. The non-negative constraints on lower-rank factorized matrices of NMF (i.e. document-topic and topic-term matrices) is particularly useful in learning topic models from text data, as all elements of document-term matrices constructed from large text corpus are naturally non-negative. Also, unlike latent semantic structures/indexing based on SVD, the latent semantic space derived by NMF does not have to be orthogonal [18], which means that NMF can learn latent semantic directions for different topics but allow overlapping terms over multiple topics.

Another important issue of learning topic models based on typical NMF is that, although a topic can be considered an additive representation of multiple terms with weights/proportions from its coefficient matrix, there is no way to control the sparseness of the coefficient matrix (i.e. the degree of how "active" of these terms are) [8]. It suggests that we may obtain a topic consisting of some key terms along with hundreds of trivial terms. Researchers have proposed various sparseness-constrained NMF learning algorithms that tackle this problem by imposing sparseness constraints on basis matrix and/or coefficient matrix in order to learn more local and prominent features/patterns [8], [19]. In this paper, we consider using *Nonsmooth Nonnegative Matrix Factorization* (*nsNMF*) [20] that puts sparseness constraints on both basis and coefficient matrices so as to extract highly localized patterns and better interpretability of the roles of documents and terms in topics. That is, for example, suppose we have identified and named multiple abstract topics (e.g. "Finance", "Politics", and "Science") from thousands of news articles using NMF. And we found that these topic share the same term "unemployment rate". However, what is the proportion of "unemployment rate" in each topics? Also, what is the proportion of all documents associated with a particular topic? Both new sparse basis and coefficient matrices created by nsNMF may answer these questions.

Besides, since the advent of topic modeling techniques, text analytics communities are interested in monitoring the development of known topics and discovering the emergences of new topics, as text data usually accumulate over time. Also, the topics must be dynamically adapted to new text data so that the new trends can be captured by the modeling algorithms. To cope with this problem, dynamic and evolutionary topic modeling techniques are proposed in the recent years. Evolutionary Nonnegative Matrix Factorization [21] aims at saving space cost and computational time by incrementally updating factorized matrices without re-running the whole NMF learning procedure at each time stamp, while a dynamic version of NMF with temporal regularization [6] learns the development of topics by preventing existing topic members from significant drifts (as evolving topics) and allowing for insertion of new small topics into the model (as emerging topics) for detection purpose. In this paper, however, we do not consider such online cases of the topic models. Instead, we define the discovery of topic evolution as the detection of whether the proportions/probabilities of given terms in all topics have changed significantly. In the next section, we discuss our proposed approach in detail.

## 3. Topic Diffusion Discovery based on Sparseness-constrained NMF

Consider that we have a large text corpus with $n$ articles/documents and a well-defined domain-specific dictionary with $p$ terms. We can create a non-negative document-term matrix $X$ where each element contains the frequency of a term that occurs in a document. Let $X$ be the $n$ by $p$ document-term matrix with non-negative frequency values. NMF is used to find two non-negative matrices $W$ and $H$ such that

$$X \approx WH$$

where $W$ is an $n$ by $k$ *basis* matrix and $H$ is a $k$ by $p$ *coefficient* matrix. The goal of a typical NMF is to find a low-rank approximation to a data matrix $X$ by minimizing the Frobenius norm $\|.\|_F$, as:

$$\min_{W,H} f(W,H) \equiv \frac{1}{2}\|X-WH\|_F^2, s.t. W \geq 0, H \geq 0$$

where $k < \min(n, p)$ and all elements in $W$ and $H$ are also non-negative. $W$ and $H$ are considered lower dimensional representations of original $X$ matrix in $k$-dimensional space. As discussed previously, NMF has been applied to various fields to learn parts of interest from data matrices. Also, imposing sparseness constraints on $W$ and $H$ can extract more localized (less-overlapping) features/patterns of the data, which may improve the interpretability of $W$ and $H$. In this paper, we propose using a normalized Nonsmooth Nonnegative Matrix Factorization (nsNMF) [20], which is originally defined as:

$$X \approx WSH$$

where $S \in \mathbb{R}^{k \times k}$ is a positive symmetric smoothing matrix defined as:

$$S = (1-\theta)I + \frac{\theta}{k}11^T$$

where $I$ is the identity matrix, $1$ is a vector of ones and $k$ is the factorization rank, and $\theta$ is the smooth parameter between 0 and 1 used to control the sparseness of $W$ and $H$. $W$ and $H$ can be considered a topic-wise document representation and term-wise topic representation, respectively. Due to the sparseness constraints on $W$ and $H$, a topic is composed of high-weighted terms, whereas a document is associated with fewer major topics. To detect whether the proportions of a given term in the topics have changed significantly, however, we need to know "what the proportion of a term is in a topic" and "what the proportion of a topic is in a document", which can be answered by row-normalizing $H$. We can do it by rewriting the original nsNMF as:

$$X \approx WSH = WSD_h D_h^{-1} H = W(SD_h)(D_h^{-1}H) = W\hat{S}\hat{H}$$

where $D_h^{-1}$ is an inverse scaling diagonal matrix with corresponding row sums of $H$ used to normalize $H$. $\hat{S}$ and $\hat{H}$ are new smoothing matrix and normalized $H$, respectively. Figure 1 shows how a document-term matrix $X$ is factorized by proposed normalized nsNMF.

The normalized nsNMF can therefore produce the output similar to those of other probabilistic topic models

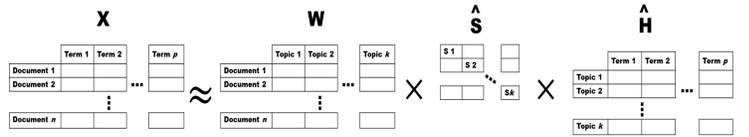

**Figure 1: Normalized nsNMF**

(e.g. LDA) but more localized and interpretable patterns. In $\hat{H}$, the proportion of a term in a topic can also be considered the conditional probability of a term $i$ given a topic $k$ (i.e. $P(term_i|topic_k)$ and $\sum_{i=1}^{p}\hat{H}_{ki}=1$). As discussed, instead, our goal is to monitor the changes of probability distributions of topics associated with a given term over time, which is to continuously compute the conditional probability of topic $k$ given a term $i$ $P(topic_k|term_i)$, and evaluate the diffusions among them in different times. The conditional probabilities $P(topic_k|term_i)$ can be obtained by Bayes' rule, as:

$$P(topic_k|term_i) = \frac{P(term_i|topic_k)P(topic_k)}{P(term_i)}$$

where $P(term_i|topic_k)$ is the proportion of a term in a topic obtained from $\hat{H}$. Also, the probability of a topic $k$ $P(topic_k)$ is the percentage of all documents associated with topic $k$ in $W$, whereas the probability of a term $i$ $P(term_i)$ is the percentage of all topics associated with term $i$ in $\hat{H}$. Note that, by "associated" here, we mean those elements with $P(document_n|topic_k)>0$ in $W$ and $P(term_i|topic_k)>0$ in $\hat{H}$ respectively. Next, to further evaluate the magnitude of the topic diffusions, we consider monitoring the aforementioned $D_{GJS}$ [9–10], which is here defined as:

$$D_{GJS}(P_1, P_2,..., P_t) = H\left(\sum_{i=1}^{t} \pi_i P_i\right) - \sum_{i=1}^{t} \pi_i H(P_i)$$

where $\pi_i$ is the weight for each discrete probability distribution. We consider assigning equal weight for each $P(topic_k|term_i)$ in time $t$. Also, $H(x)$ is $k$-ary Shannon entropy that is defined as:

$$H(x) = -\sum_{i=1}^{k} P(x_i) \log_k P(x_i)$$

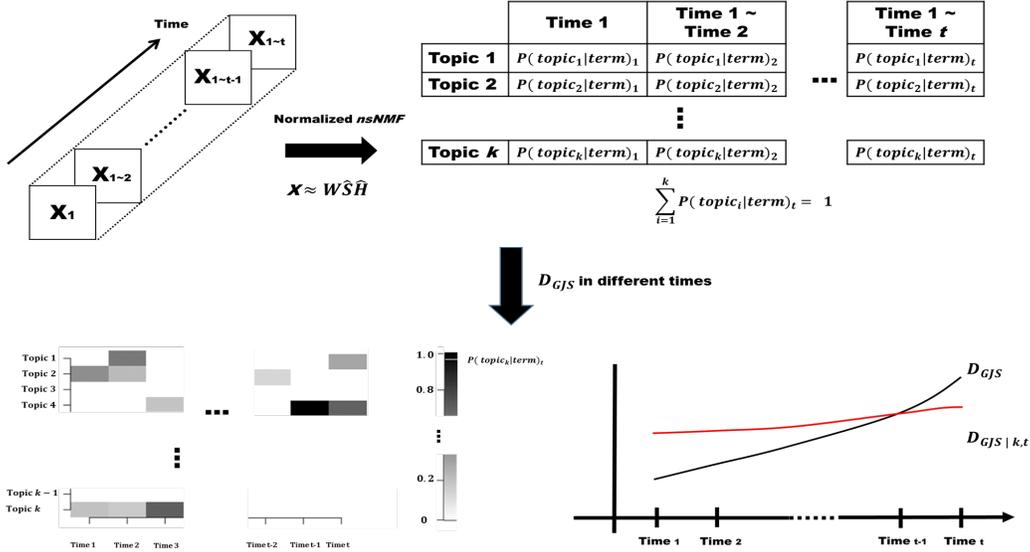

**Figure 2: Overview of Proposed Topic Diffusion Discovery Process**

The $D_{GJS}$ is a symmetric measure that ranges between 0 and 1. Also defined in [9–10], a statistically significant threshold of the generalized Jensen-Shannon divergence $D_{GJS|k,t}$ can be asymptotically approximated and represented by Chi-square statistic $\chi^2$, as

$$D_{GJS|k,t} \simeq \frac{X^2_{df, 1-\alpha}}{2N(\ln_k)}$$

where $df = (k-1)(t-1)$ is the degree of freedom, $\alpha$ is the statistical significance level (usually 0.05 or 0.01), and $N$ is the total number of cells ($k$ by $t$) used in calculating the Chi-square statistic $\chi^2$ in different times.

Also notice that, as these topics generated by NMF in different periods are not always aligned with each other due to randomness, we have to match these topics by identifying minimum "distances" among conditional probability distributions given topics in different periods. For example, suppose we have 2 terms and 3 topics ($topic_1$, $topic_2$, and $topic_3$) for 2 periods ($time_1$ and $time_2$), and discrete conditional probability distributions $P(term|topic_k)$ (0.5, 0.5), (0.1, 0.9), (0.9, 0.1) for $time_1$ and (0.8, 0.2), (0.4, 0.6), (0.1, 0.9) for $time_2$, respectively. The best topic matches should be (1→2, 2→3, 3→1), according to the distances/divergences among topics. To identify the best matches of topics over the time, we propose using Hungarian (Kuhn–Munkres) algorithm [7] with pairwise Jensen-Shannon divergences as the cost measures.

Figure 2 shows an overview of our proposed approach. We here summarize the steps of the approach that helps discover topic diffusions, which is, again, *whether the probability distributions of topics associated with a given term have changed significantly over time.* We first create document-term matrix $X_t$ based on document data and pre-defined dictionary. After applying the normalized nsNMF to the matrix $X_t$, we can obtain the conditional probabilities of $k$ topics for a given term. The $D_{GJS}$ and $D_{GJS|k,t}$ at time $t$ can later be calculated. We keep updating and re-running this process when new documents/text data appears. By checking whether $D_{GJS}$ is higher or lower than the threshold $D_{GJS|k,t}$ with $\alpha = 0.01$, we can learn whether the given term is *convergent* or *divergent*, and whether there is any significant change on the given term over the topics. Take the "Electronics" and "Lifestyle" topic evolution in Section 1 as an example, we here would like to know how "cell phone" or "camera" is associated with topic "Electronics" and "Lifestyle" over time.

## 4. Experiment and Discussion

In this section, we present our experimental results and discuss findings of the proposed approach. To evaluate feasibility of the approach, we collected open-access and publicly available articles related to "Machine Learning" in 2004/01-2016/12 from *arXiv.org stat.ML* [22].

The goal of using these articles as the text data is to evaluate whether proposed approach can extract meaningful topics and identify the relationships among research keywords/terms and topics related to Machine Learning. All the experiments were implemented in *R* 3.4.3 [23] and were performed on a computing server with two Intel Xeon CPUs and an NVIDIA Geforce GTX 1080 Ti GPU. The source code and data are available upon request. We extracted 7962 raw keywords in articles entered by article authors as the basic lexicon to build a dictionary specific to Machine Learning and related fields. These keywords were manually reviewed by domain experts, and a table of keyword processing rules were created and used to remove/correct redundant and inaccurate keywords. There are total 4187 keywords/terms in the final dictionary and 9044 articles/documents from 2004/01 to 2016/12. We used R package *tm* 0.6-2 [24] to create document-term matrices for later topic analysis after a series of text transformations, including whitespace elimination, lower case conversion, inaccurate term replacement, stopword removal, and tf-idf transformation. Note that we did not consider stemming here, as we used our pre-defined dictionary with domain-specific and discriminative terms based on the article keywords from authors.

To understand the topic diffusions in Machine Learning, we empirically chose factorization rank $k = 10$ to extract ten abstract topics using R package *nmfgpu4R* [25]. Also, as there are relative fewer articles in arxiv stat.ML in 2004-2011, we consider creating six document-term matrices for 2004-2011, 2004-2012, 2004-2013, 2004-2014, 2004-2015, and 2004-2016 in order to understand the development of terms and topics. There are 980, 2118, 3296, 4635, 6439, and 9044 accumulated articles during these periods, respectively. The proposed normalized nsNMF was then applied to six matrices to obtain the conditional probabilities for each term given a topic, $P(term|topic_k)$. Also notice that here the sparseness parameter $\theta$ is empirically set to 0.4, which could impose sparseness constraints on both basis and coefficient matrices without losing too much information in original data matrices [20].

Table 1 shows top-10 terms of the ten topics over six periods. We can see that most topics seem to exhibit clear subjects, but there are still a few topics that show some drifts and become hard to interpret. Literature has indicated that it is most likely because existing topics may evolve and new topics may emerge [26–27]. And therefore the choice of the number of topics may need to change over time. In recent years, although some heuristic methods/measures such as stability [18] and separability [27] analyses are proposed, finding the optimal number of topics has still been a key issue and an open question in topic modeling [26–27]. Too few topics may produce extremely broad topics, whereas too many topics may result in numerous narrow and similar topics. Therefore, we here instead would like to know how given terms are associated with topics as well as how topics evolve with times. In the following figures, we present and evaluate the diffusion of topics and terms based on monitoring the aforementioned $D_{GJS}$ along with tile plots that indicate the conditional probability of topic $k$ given a term $i$ $P(topic_k|term_i)$.

A narrow topic may be represented by a few terms, and a broad term may be considered a topic. To further understand the topic diffusions, we here discuss patterns of the diffusion in terms of topic/term *broadness* and *convergence* along with $D_{GJS}$. Figure 3 shows examples of broad and divergent terms. We can see that these terms, "sequence data", "spatiotemporal" and "semisupervised learning", can be considered relatively broad "topics". They are associated with more topics and their $D_{GJS}$ are getting higher over time. It suggests that these terms are widely discussed and used in many fields of machine learning in 2004-2016, and there is almost no sign of topic convergences related to these terms. On the other hand, there are indeed broad but convergent terms. We found that "reinforcement learning", "linear model", and "statistical learning", as shown in Figure 4, are associated with a few topics and their $D_{GJS}$ are lower than the thresholds given $\alpha = 0.01$, which indicates that they are convergent terms and are only used in some fields.

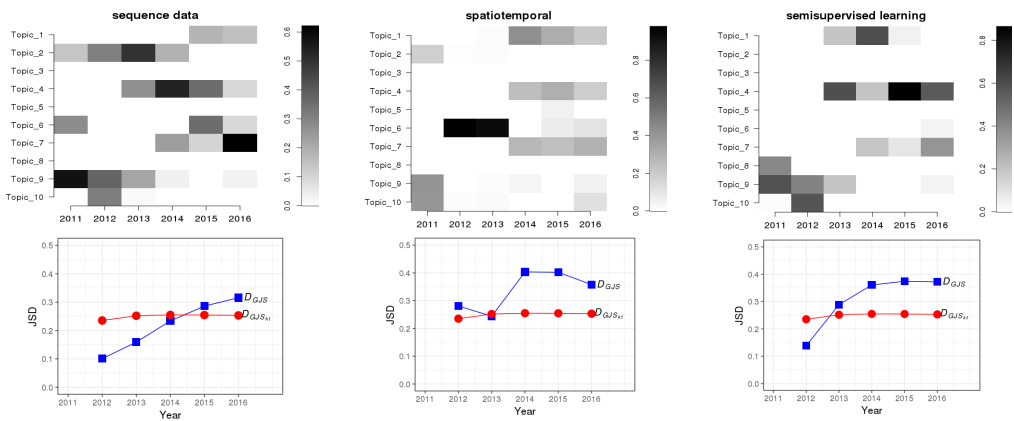

**Figure 3: Broad and divergent terms**

| | ~2011 | ~2012 | ~2013 | ~2014 | ~2015 | ~2016 |
|---|---|---|---|---|---|---|
| **Topic 1** | dendrogram<br>ultrametric<br>clustering<br>hierarchical clustering<br>hierarchical<br>agglomerative<br>topology<br>hierarchy<br>wavelet<br>baire distance | dendrogram<br>ultrametric<br>clustering<br>hierarchical clustering<br>hierarchical<br>agglomerative<br>baire distance<br>hierarchy<br>dissimilarity<br>topology | clustering<br>dendrogram<br>ultrametric<br>hierarchical clustering<br>spectral clustering<br>hierarchical<br>agglomerative<br>similarity<br>dissimilarity<br>hierarchy | clustering<br>spectral clustering<br>dendrogram<br>ultrametric<br>hierarchical clustering<br>hierarchical<br>similarity<br>agglomerative<br>graph<br>dissimilarity | clustering<br>spectral clustering<br>hierarchical clustering<br>dendrogram<br>subspace clustering<br>dissimilarity<br>hierarchical<br>similarity<br>linkage<br>agglomerative | clustering<br>hierarchical clustering<br>spectral clustering<br>subspace clustering<br>dendrogram<br>hierarchical<br>ultrametric<br>similarity<br>dissimilarity<br>agglomerative |
| **Topic 2** | markov equivalence<br>graph<br>markov<br>hidden markov model<br>network<br>independence<br>mixed graph<br>graphical<br>independence model<br>graphical model | topic<br>dirichlet<br>latent<br>latent dirichlet allocation<br>topic models<br>dirichlet process<br>variational<br>perplexity<br>mixture<br>posterior | topic<br>dirichlet<br>latent<br>latent dirichlet allocation<br>variational<br>posterior<br>topic models<br>markov<br>mixture<br>inference | topic<br>dirichlet<br>latent<br>latent dirichlet allocation<br>topic models<br>variational<br>dirichlet process<br>perplexity<br>vocabulary<br>news | topic<br>dirichlet<br>latent<br>latent dirichlet allocation<br>topic models<br>perplexity<br>dirichlet process<br>variational<br>vocabulary<br>topic modeling | topic<br>dirichlet<br>latent dirichlet allocation<br>latent<br>topic models<br>perplexity<br>topic modeling<br>dirichlet process<br>news<br>vocabulary |
| **Topic 3** | regret<br>bandit<br>ucb<br>approachability<br>online<br>games<br>regret bounds<br>hedge<br>minimax<br>calibration | regret<br>bandit<br>ucb<br>regret bounds<br>online<br>games<br>minimax<br>bandit problems<br>feedback<br>contextual bandit | regret<br>bandit<br>ucb<br>online<br>regret bounds<br>feedback<br>hedge<br>online learning<br>games<br>contextual bandit | regret<br>bandit<br>ucb<br>online<br>regret bounds<br>feedback<br>online learning<br>hedge<br>games<br>adversarial | regret<br>bandit<br>ucb<br>online<br>regret bounds<br>feedback<br>contextual bandit<br>online learning<br>minimax<br>adversarial | regret<br>bandit<br>ucb<br>online<br>regret bounds<br>feedback<br>contextual bandit<br>online learning<br>optimal<br>minimax |
| **Topic 4** | meld<br>survival<br>survival tree<br>recursive partitioning<br>clinical<br>survival analysis<br>conditional inference trees<br>permutation<br>trees<br>covariate | meld<br>survival<br>survival tree<br>recursive partitioning<br>clinical<br>survival analysis<br>conditional inference trees<br>permutation<br>trees<br>permutation tests | graph<br>network<br>markov<br>markov equivalence<br>graphical<br>bayesian network<br>independence<br>graphical model<br>bayesian<br>independence model | graph<br>markov<br>network<br>bayesian<br>posterior<br>markov chain<br>hidden markov model<br>graphical<br>bayesian network<br>monte carlo | graph<br>network<br>community<br>block model<br>detection<br>community detection<br>modularity<br>graphical<br>bayesian network<br>stochastic block model | graph<br>network<br>community<br>detection<br>laplacian<br>diffusion<br>community detection<br>markov equivalence<br>block model<br>graphical |
| **Topic 5** | sar<br>sar image<br>atr<br>scattering<br>em simulator<br>bistatic radar<br>simulation<br>synthetic aperture radar<br>principal component analysis<br>component analysis | sar<br>sar image<br>atr<br>scattering<br>em simulator<br>speckle<br>stochastic distances<br>bistatic radar<br>polarimetric sar<br>simulation | sar<br>sar image<br>speckle<br>scattering<br>atr<br>stochastic distances<br>em simulator<br>polarimetric sar<br>synthetic aperture radar<br>remote sensing | sar<br>sar image<br>speckle<br>scattering<br>atr<br>stochastic distances<br>em simulator<br>polarimetric sar<br>synthetic aperture radar<br>remote sensing | diffusion<br>manifolds<br>diffusion maps<br>embedding<br>geodesic<br>dictionary learning<br>laplacian<br>vector diffusion maps<br>heat kernel<br>component analysis | gaia<br>quasars<br>photometric systems<br>contamination<br>extinction<br>completeness<br>mutation<br>degeneracy<br>evolutionary algorithms<br>galaxies |
| **Topic 6** | quasars<br>galaxies<br>redshift<br>gaia<br>contamination<br>extinction<br>completeness<br>photometric systems<br>degeneracy<br>galaxy clusters | searchlight<br>mvpa<br>cubes<br>member<br>separation<br>fmri<br>monotonic<br>human<br>decoding<br>multivariate | searchlight<br>mvpa<br>cubes<br>member<br>monotonic<br>fmri<br>separation<br>human<br>decoding<br>classification accuracy | beta process<br>completely random measure<br>nonparametric<br>gamma process<br>dirichlet process<br>indian buffet process<br>bayesian nonparametric<br>point process<br>predictive distribution<br>factor analysis | markov<br>posterior<br>markov chain<br>variational<br>monte carlo<br>hidden markov model<br>bayesian<br>markov chain monte carlo<br>inference<br>particle | variational<br>markov<br>posterior<br>markov chain<br>monte carlo<br>bayesian<br>markov chain monte carlo<br>latent<br>inference<br>mixture |
| **Topic 7** | distortion measure<br>distortion<br>observations<br>self organizing map<br>law of large numbers<br>strong<br>phase transition<br>assessment<br>free energy<br>minimization | kriging<br>covariance<br>additive<br>interpretation<br>reliability<br>random process<br>data assimilation<br>expected improvement<br>observations<br>gam | omp<br>matching pursuit<br>orthogonal matching pursuit<br>matching<br>sparse<br>rip<br>greedy algorithm<br>sparse recovery<br>measurements<br>restricted isometry | support vector<br>support vector machine<br>classifier<br>training<br>classification<br>margin<br>ensemble<br>boosting<br>adaboost<br>ranking | training<br>classifier<br>support vector<br>ranking<br>rank<br>classification<br>component analysis<br>convex<br>support vector machine<br>dropout | recurrent neural networks<br>neural networks<br>lstm<br>dropout<br>adversarial<br>training<br>network<br>autoencoder<br>deep learning<br>machine translation |
| **Topic 8** | adaboost<br>boosting<br>margin<br>classifier<br>logit<br>training<br>ensemble<br>convergence<br>differential equation<br>rate of convergence | smml<br>minimum message length<br>exponential family<br>exponential<br>rules<br>mml<br>local minimum<br>statistical models<br>step functions<br>kolmogorov complexity | smml<br>exponential family<br>minimum message length<br>exponential<br>mml<br>partition<br>rules<br>data compression<br>step functions<br>kolmogorov complexity | smml<br>exponential family<br>minimum message length<br>exponential<br>mml<br>partition<br>rules<br>data compression<br>step functions<br>kolmogorov complexity | smml<br>exponential family<br>minimum message length<br>exponential<br>mml<br>partition<br>rules<br>data compression<br>step functions<br>kolmogorov complexity | smml<br>mml<br>minimum message length<br>exponential family<br>exponential<br>ideal group<br>partition<br>data compression<br>kolmogorov complexity<br>rules |
| **Topic 9** | kernel<br>reproducing kernel<br>support vector<br>hilbert space<br>support vector machine<br>reproducing kernel hilbert<br>density<br>component analysis<br>minwise hashing<br>regularization | kernel<br>reproducing kernel<br>hilbert space<br>support vector<br>reproducing kernel hilbert<br>support vector machine<br>margin<br>adaboost<br>kernel learning<br>boosting | kernel<br>reproducing kernel<br>support vector<br>hilbert space<br>reproducing kernel hilbert<br>support vector machine<br>kernel learning<br>training<br>classification<br>density | kernel<br>reproducing kernel<br>hilbert space<br>reproducing kernel hilbert<br>kernel learning<br>kernel matrix<br>embedding<br>positive definite<br>feature map<br>gaussian kernel | kernel<br>reproducing kernel<br>hilbert space<br>reproducing kernel hilbert<br>support vector<br>kernel learning<br>support vector machine<br>kernel matrix<br>positive definite<br>feature map | kernel<br>reproducing kernel<br>hilbert space<br>reproducing kernel hilbert<br>support vector<br>support vector machine<br>kernel matrix<br>feature map<br>embedding<br>kernel learning |
| **Topic 10** | lasso<br>group lasso<br>sparsity<br>sparse<br>adaptive lasso<br>regression<br>glasso<br>selection<br>variable selection<br>penalized | graph<br>lasso<br>markov<br>network<br>sparse<br>graphical<br>group lasso<br>bayesian<br>gene<br>covariance | lasso<br>group lasso<br>sparse<br>sparsity<br>regression<br>selection<br>convex<br>screening<br>norm<br>covariance | lasso<br>group lasso<br>screening<br>sparse<br>sparsity<br>regression<br>selection<br>penalized<br>variable selection<br>adaptive lasso | lasso<br>screening<br>group lasso<br>sparsity<br>sparse<br>selection<br>regression<br>variable selection<br>penalized<br>elastic net | lasso<br>sparse<br>convex<br>screening<br>rank<br>component analysis<br>sparsity<br>norm<br>regression<br>principal component analysis |

**Table 1: Top-10 terms over 10 topics for six periods**

We next would like to know whether the proposed approach could also help discover topic diffusions given relatively narrow terms. In Figure 5 and Figure 6, we consider those terms/techniques used in various model learning tasks. Figure 5 shows that these techniques can be regarded as divergent terms because of the upward trends of the $D_{GJS}$. We may argue that they have been applied to various fields of machine learning in the past years. On the contrary, terms in Figure 6 should be considered convergent. For example, although "least squares" is commonly used in learning parameters (weights) of linear models, but its applications are relatively limited compared to other techniques such as "gradient descent". Also, "markov decision process" and "link prediction" are popular terms in reinforcement learning and network data analysis, but we may say that they has not been widely applied to other fields of machine learning.

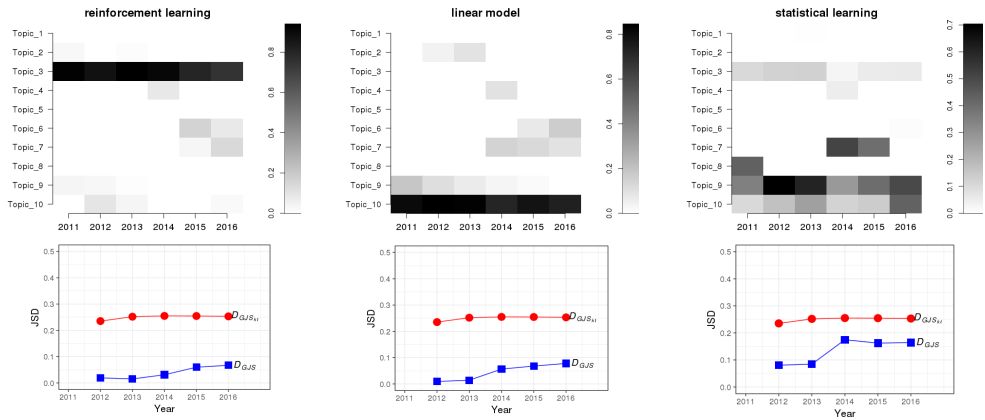

**Figure 4: Broad but convergent terms**

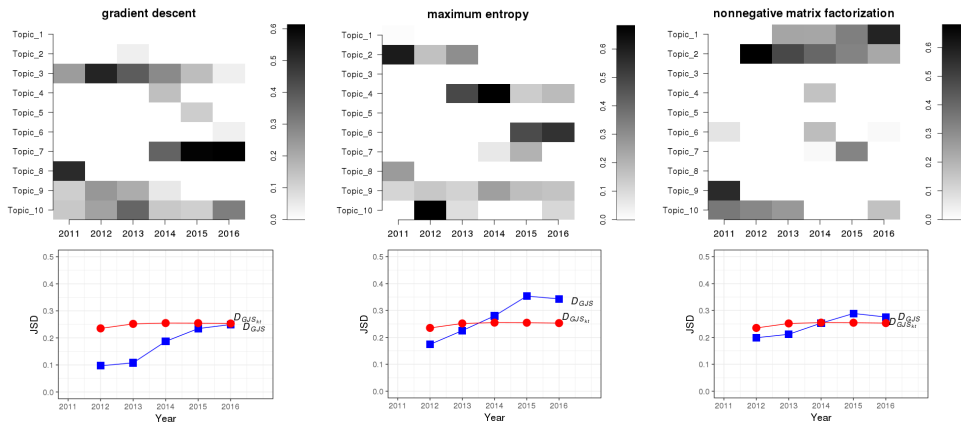

**Figure 5: Narrow but divergent terms**

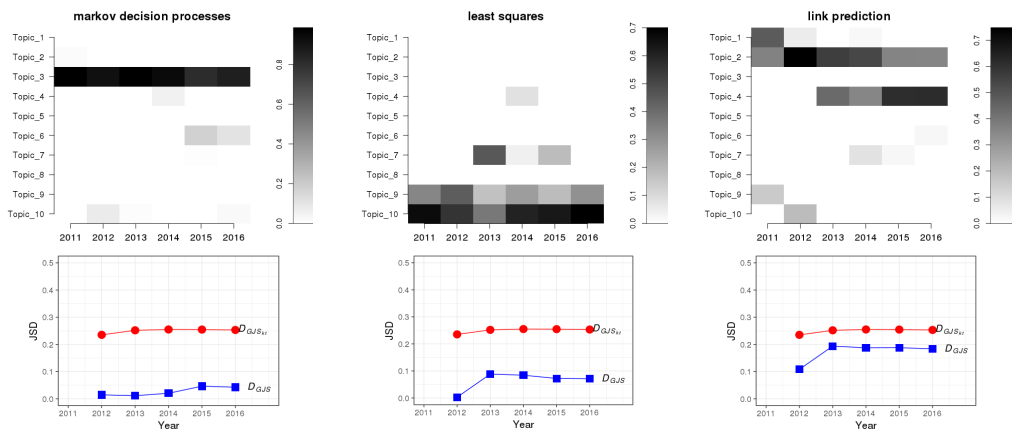

**Figure 6: Narrow and convergent terms**

## 5. Conclusion

We proposed a novel approach that combines non-negative matrix factorization with generalized Jensen-Shannon divergence to discover topic diffusions from large text/document data. The experiment results show that it can help build topic models with less-overlapping topics, uncover connections between terms and topics, identify convergent/divergent terms, and determine whether topics/terms have been widely used in literatures. The approach is applied to a large number of research articles and demonstrated to be able to help researchers understand the development of research topics and discover the diffusion of specific terms.

## 6. References

[1] M. Steyvers and T. Griffiths, "Probabilistic topic models," *Handbook of latent semantic analysis*, vol. 427, no. 7, pp. 424–440, 2007.

[2] D. M. Blei, A. Y. Ng, and M. I. Jordan, "Latent dirichlet allocation," *Journal of machine Learning research*, vol. 3, no. Jan, pp. 993–1022, 2003.

[3] D. D. Lee and H. S. Seung, "Learning the parts of objects by non-negative matrix factorization," *Nature*, vol. 401, no. 6755, pp. 788–791, 1999.

[4] L. AlSumait, D. Barbará, and C. Domeniconi, "On-line LDA: Adaptive topic models for mining text streams with applications to topic detection and tracking," in *2008 eighth IEEE international conference on data mining*, 2008, pp. 3–12.

[5] M. Hoffman, F. R. Bach, and D. M. Blei, "Online learning for latent dirichlet allocation," in *advances in neural information processing systems*, 2010, pp. 856–864.

[6] A. Saha and V. Sindhwani, "Learning evolving and emerging topics in social media: a dynamic NMF approach with temporal regularization," in *Proceedings of the fifth ACM international conference on Web search and data mining*, 2012, pp. 693–702.

[7] J. Choo, C. Lee, C. K. Reddy, and H. Park, "Utopian: User-driven topic modeling based on interactive nonnegative matrix factorization," *IEEE transactions on visualization and computer graphics*, vol. 19, no. 12, pp. 1992–2001, 2013.

[8] P. O. Hoyer, "Non-negative matrix factorization with sparseness constraints," *Journal of machine learning research*, vol. 5, no. Nov, pp. 1457–1469, 2004.

[9] Y. Kang and V. Zadorozhny, "Process Monitoring Using Maximum Sequence Divergence," *Knowledge and Information Systems*, vol. 48, no. 1, pp. 81–109, Jul. 2016.

[10] I. Grosse, P. Bernaola-Galván, P. Carpena, R. Román-Roldán, J. Oliver, and H. E. Stanley, "Analysis of symbolic sequences using the Jensen-Shannon divergence," *Physical Review E*, vol. 65, no. 4, p. 041905, 2002.

[11] C. C. Aggarwal and C. Zhai, *Mining text data*. Springer Science & Business Media, 2012.

[12] D. M. Blei, "Probabilistic topic models," *Communications of the ACM*, vol. 55, no. 4, pp. 77–84, 2012.

[13] D. M. Blei and J. D. Lafferty, "Dynamic topic models," in *Proceedings of the 23rd international conference on Machine learning*, 2006, pp. 113–120.

[14] L. Gong and A. K. Nandi, "An enhanced initialization method for non-negative matrix factorization," in *Machine Learning for Signal Processing (MLSP), 2013 IEEE International Workshop on*, 2013, pp. 1–6.

[15] T. K. Landauer, P. W. Foltz, and D. Laham, "An introduction to latent semantic analysis," *Discourse processes*, vol. 25, no. 2–3, pp. 259–284, 1998.

[16] P. Smaragdis and J. C. Brown, "Non-negative matrix factorization for polyphonic music transcription," in *Applications of Signal Processing to Audio and Acoustics, 2003 IEEE Workshop on.*, 2003, pp. 177–180.

[17] D. Guillamet and J. Vitria, "Non-negative matrix factorization for face recognition," in *Topics in artificial intelligence*, Springer, 2002, pp. 336–344.

[18] S. Arora, R. Ge, and A. Moitra, "Learning topic models–going beyond SVD," in *Foundations of Computer Science (FOCS), 2012 IEEE 53rd Annual Symposium on*, 2012, pp. 1–10.

[19] H. Kim and H. Park, "Sparse non-negative matrix factorizations via alternating non-negativity-constrained least squares for microarray data analysis," *Bioinformatics*, vol. 23, no. 12, pp. 1495–1502, 2007.

[20] A. Pascual-Montano, J. M. Carazo, K. Kochi, D. Lehmann, and R. D. Pascual-Marqui, "Nonsmooth nonnegative matrix factorization," *IEEE transactions on pattern analysis and machine intelligence*, vol. 28, no. 3, pp. 403–415, 2006.

[21] F. Wang, H. Tong, and C.-Y. Lin, "Towards Evolutionary Nonnegative Matrix Factorization.," in *AAAI*, 2011, vol. 11, pp. 501–506.

[22] "arXiv.org e-Print archive." https://arxiv.org/

[23] R Core Team, "R: A Language and Environment for Statistical Computing," *R Foundation for Statistical Computing, Vienna, Austria*, 2016.

[24] D. Meyer, K. Hornik, and I. Feinerer, "Text mining infrastructure in R," *Journal of statistical software*, vol. 25, no. 5, pp. 1–54, 2008.

[25] S. Koitka and C. M. Friedrich, "nmfgpu4R: GPU-Accelerated Computation of the Non-Negative Matrix Factorization (NMF) Using CUDA Capable Hardware," *The R Journal*, vol. 8, no. 2, pp. 382–392, 2016.

[26] K. Stevens, P. Kegelmeyer, D. Andrzejewski, and D. Buttler, "Exploring topic coherence over many models and many topics," in *Proceedings of the 2012 Joint Conference on Empirical Methods in Natural Language Processing and Computational Natural Language Learning*, 2012, pp. 952–961.

[27] D. Greene, D. O'Callaghan, and P. Cunningham, "How many topics? stability analysis for topic models," in *Joint European Conference on Machine Learning and Knowledge Discovery in Databases*, 2014, pp. 498–513.